\documentclass[11pt]{article}
\title{Pendulum Integration and Elliptic Functions}
\author{\small\textsc{Pedro L. Garrido}
\\
\small Institute Carlos I for \\
\small Computational and \\
\small Theoretical Physics, \\
\small Universidad de Granada, \\
\small Espa$\tilde {\rm n}$a
\and\small\textsc{Giovanni Gallavotti}
\\
\small Dipartimento di Fisica and INFN,\\ 
\small Universit{\'a} di Roma "La Sapienza",\\ 
\small Italia}%
\let\a=\alpha \let\b=\beta  \let\g=\gamma  \let\d=\delta 
     \let\th=\theta \let\k=\kappa \let\l=\lambda
        \let\x=\xi     \let\p=\pi    
    \let\f=\varphi

\let\0=\noindent
\def\defi{\,{\buildrel def\over=}\,}

\def\UU{{\cal U}}
\def\otto{\,{\kern-1.truept\leftarrow\kern-5.truept\to\kern-1.truept}\,}
\def\txt{\textstyle}
\def\EJ{{\bf E}}\def\KJ{{\bf K}}\def\dn{{\,{\rm dn}\,}}\def\sn{{\,{\rm sn}\,}}
\def\cn{{\,{\rm cn}\,}}\def\am{{\,{\rm am}\,}}
\def\*{\vskip2mm}
\def\Eq#1{{\label{#1}}%
\write15{\string\Fe{\string#1}{\ref{#1}}}}
\def\equ#1{\ref{#1}}

\def\be{\begin{equation}}\def\ee{\end{equation}}
\def\iniz{\setcounter{equation}{0}}
\renewcommand{\theequation}{\arabic{section}.\arabic{equation}}
\usepackage{eqalignno}

\begin{document}
\maketitle

\begin{abstract}\0Revisiting canonical integration of the classical
    pendulum around its unstable equilibrium, normal hyperbolic canonical
    coordinates are constructed.\end{abstract} \*

\0{\small {\bf Key words:} \it Elliptic
    Functions, Pendulum, Canonical Integrability}%
\footnote[1]{\small\texttt{giovanni.gallavotti@roma1.infn.it},
\texttt{garrido@onsager.ugr.es}
}%
\*

\section{Pendulum near the separatrix}
\iniz

The theory of Jacobian elliptic functions, for reference see
\cite{GR965}, yields a complete calculation for the motion of a
pendulum as a function of time.  This is revisited here, to exhibit a
few interesting properties of the elliptic integrals.

Write the pendulum energy, with inertia moment $I$ and gravity
constant $g^2$ (rather than the usual $g$), in the canonical coordinates
$(B,\b)$ or as
\be {B^2\over2I}-I g^2(1-\cos\b)\defi H(B,\b)\Eq{e1.1}\ee
where the origin in $\b$ is set at the unstable equilibrium: the
definition implies that $g$ has dimension of inverse time and the
Lyapunov exponents of the unstable equilibrium are $\pm g$. $B\defi
I\dot\b$ and $\b$ are canonical coordinates for the motions.

It is well known that near the unstable equilibrium of the pendulum
$B=0,\b=0$ it is possible to define a canonical transformation,
mapping the origin into itself, introducing new local cordinates
$(p,q)$ such that

\be B=R_c(p,q),\quad \b=S_c(p,q)\Eq{e1.2}\ee
with $R,S$ holomorphic in a polidisk $|p|,|q|<\k$ with $\k>0$, and in
terms of which the motion near $B=\b=0$ is described by a Hamiltonian
$G$ depending on the product $pq$ only, of the form $\UU(p\cdot
q)=H(B,\b)$ with $\frac{d\,\UU}{d(pq)}(0)=g$.

The purpose of this note is to derive a proof of the latter statement
via the theory of elliptic functions: this is not the simplest
approach if one is just interested to know the existence of normal
hyperbolic coordinates: existence of $R,S$ could be easily established
without deriving their ``explicit'' expressions for $p$ and $q$ in
terms of elliptic functions. Here we also correct a few errors in the
earlier attempt made in \cite[Appendix 9]{CG994}.

The natural correspondence between the hyperbolic fixed point of the
pendulum and its elliptic fixed point is briefly reported in Appendix
B and leads to the construction of the normal canonical coordinates for
the small oscillations, hence to the action-angle variables.

\section{Solution in terms of elliptic integrals}
\iniz

Motions near the unstable equilibrium have a quite different nature
depending on the sign of the total energy $H(B,\b)=U$: the ones with
$U<0$ are ``oscillations'' (their motions do not encompass the full
perimeter of the circle) while the ones with $U>0$ are
``librations''. Therefore it will not be possible to find global
action-angle coordinates: motions near the separatrix (which with our
conventions has $U=0$) require other coordinates to be expressed in a
simple way.

Introduce the variables that appear in the theory of
Jacobi's elliptic functions

\be
\eqalign{
k'\,=\,&\sqrt{1-k^2},\qquad
h'\,=\,\frac{k}{\sqrt{1+k^2}},\qquad {h}\,=\,\sqrt{1-{h'}^2}
\cr
U\,=\,& 2 g^2 I\frac{1}{k^2},\qquad \kern2mm
u\,=\,t\,\sqrt{\frac{U}{2I}}\qquad\kern0.5cm{g}_0\,=\,
g\frac\p{2\,h'\, \KJ(h)}
\cr}\Eq{e2.1}\ee
where, ${\bf K}(k)\defi
\int_0^{\frac\p2}{(1-k^2\sin^2\a)}^{-\frac{1}2}d\a$. Hence the
separatrix has $k=+\infty$ and $U=0$; and the data {\it above the
separatrix} correspond to $U>0$ (or $k>0$). Note that $g_0(0)=g$
because ${\bf K}(0)=\frac\p2$ (as $U=0$ corresponds to $k=\infty$ and
$h'=1,h=0$); the following formulae become singular as $U\to0$, but
the singularity is only apparent and it will disappear from all
relevant formulae derived or used in the following.

Other important quantities in the elliptic functions theory are, see
the references
\cite[(8.198.1)]{GR965},\cite[(8.198.2),(8.146)]{GR965},
\be \eqalign{
x'\defi\x(h)=&e^{-\p\KJ(h')/\KJ(h)}=\l+2\l^5+15\l^9+..\cr
\l\defi&{1\over2}\frac{1-\sqrt{h'}}{1+\sqrt{h'}}= \frac{\sum_{n=0}^\infty
\x(h)^{(2n+1)^2}}{1+2 \sum_{n=1}^\infty \x(h)^{4n^2}}\cr
}\Eq{e2.2}\ee
where $\x(k')$ denotes here what in \cite{GR965} would be $q(k')$
(\cite[(8.146.1),(8.194.2)]{GR965}).

In terms of the above conventions we have, directly from the
definitions of $\am,\cn,\sn,\dn$ (Jacobi's elliptic functions,
\cite[(8.14)]{GR965}), and from the equations of motion:
\be
\b(t)=2\am(u,ik),\, u=\frac{tg}{k},\qquad
B(t)=I\dot \b=\frac{2Ig}k\dn(u,ik)\Eq{e2.3}\ee
(\cite[(8.143),(8.141)]{GR965}). So that the action $B$ is given as a
function of time
\be B(t)=\frac{2Ig }{k\,\dn(\frac{u}{h},h')}
=2Ig \, \frac{{\rm cn}(-i\frac{u}{h},h)}
{k\,\dn(-i\frac{u}{h},h)},\Eq{e2.4}\ee
(\cite[(8.153.9),(8.153.3)]{GR965}) assuming that initial data are
assigned with $\b=0$.

The $t$ dependence of $B(t)$ is naturally expressed via the argument
$\frac{u}h=\frac{g t}{kh{\bf K}(h)}$, if the second of Eq.(\equ{e2.4})
is used, since $kh\equiv h'$, see Eq.(\equ{e2.1}). This explains the
role that

\be g_0(x')
\defi g_0\equiv \frac\p2\frac{g}{kh{\bf K}(h)}=\frac\p2
\frac{g}{h'\KJ(h)}\Eq{e2.5}\ee
Eq.(\equ{e2.5}), will play in the following analysis an important
role, and it admits a rather simple product expansion,
 \cite[(8.197.1),(8.197.4)]{GR965},

\be
g_0(x')=g \prod_{n=1}^{\infty}\Big(\frac{1+x^{'n}}{1-x^{'n}}\Big)^2\Eq{e2.6}
\ee
and its logarithmic derivative is $4\sum_{n=1}^\infty \frac{n
  {x'}^{n-1}}{1-{x'}^{2n}}$ so that $x'\frac{d}{dx'}\log g_0(x')$ is
  $\frac12\frac{d^2}{d{z}^2} \log \th_4(z,x')\Big|_{z=0}$, where
  $\th_4(z,x')$, \cite[p.463,489]{WW927}, is a Jacobi's theta function.

It is also convenient to remark that in a motion with energy $U$ it
will be

\be H(B(t),\b(t))\equiv U=2 g^2 I\frac{1}{k^2}\Eq{e2.7}\ee

\section{Power series representation}
\iniz

From the theory of elliptic functions the evolution $B(t),\b(t)$ with
any initial data {\it above the separatrix} ({\it i.e.}  with
$\b(0)=0$ and $B(0)=I\dot\b(0)$ corresponding to a given value of $h$,
with $U>0$), can be expressed as $B(t)=\overline R(\g,\d)$ and
$\b(t)=\overline S(\g,\d)$ with $\g = e^{g_0 t}, \d= e^{-g_0t}$ and,
taking into account \cite[(8.146.11)]{GR965},
\be \eqalign{
\txt \overline R(\g,\d)=-4g_0 I\mathop\sum\limits_{n=1}^\infty
\Bigl[{(-1)^n\,
\x^{n-\frac12}\,(\g^{2n-1}+\d^{2n-1})\over
1-\x^{2n-1}}\Bigr]\cr}
\Eq{e3.1}\ee
with $\x\equiv\x(h)$.  Definitions in Eq. (\equ{e2.1}),(\equ{e2.2})
yield
\be g_0(\x)=g\frac\p2
\frac1{\sqrt{1-h^2}\,\KJ(h)}=g\,(1+\frac14 h^2+\ldots)\Eq{e3.2}\ee
which is analytic in $h^2$ by \cite[(8.113.1)]{GR965} near $h=0$.

Eq.(\equ{e2.2}) implies that $\x=\l+O(\l^5)$ is analytic in $\l$
near $\l=0$ so that $h^2=16\l+\ldots=16\x+\ldots$.  Therefore
\be g_0=(1+4\x+12\x^2+\ldots)\,g\Eq{e3.3}\ee
is analytic in $\x$ near $\x=0$.

The evolution of $\f$ is then a consequence of Eq.(\equ{e3.1}) which
leads to an expression for $\overline S$ by the remark that $\overline
R=g_0I(\g\partial_\g-\d\partial_\d)\overline S$ (just expressing that $B$ is $I$
times the derivative of $\b$): namely

\be \overline S(\g,\d)=-4\mathop\sum\limits_{n=1}^\infty\frac{(-1)^n
\x^{n-\frac12}}{1-\x^{2n-1}}
\frac{\g^{2n-1}-\d^{2n-1}}{2n-1}\Eq{e3.4}\ee
and, after developing in powers of $\x$ the denominators and resumming,

\be
\eqalign{
\txt \overline S=&\txt4\mathop\sum\limits_{m=0}^\infty
\big(\arctan(\x^m \,\d\sqrt\x)-\arctan(\x^m\,\g\sqrt\x)\big)
\cr
\txt \overline R=&\txt 4I\,g_0\mathop\sum\limits_{m=0}^\infty
\big(\frac{\x^m \g\sqrt{\x}}{1+(\x^{m} \,\g\,\sqrt\x)^2}+
\frac{\x^m \d\sqrt{\x}}{1+(\x^{m}\, \d \,\sqrt\x)^2}\big)\cr}
\Eq{e3.5}\ee
The first formula reminds of one found by Jacobi which he commented by
saying that `` {\it inter formulas elegantissimas censeri debet} '',
\cite[p.509]{WW927} ({\it i.e.} `` {\it it should be counted among the
most elegant formulae} ``.

Note that $g_0$ depends only on $\x$, see Eq.(\equ{e3.2}), which would
be surprising if the mechanical interpretation was not taken into
account.  The Eq.(\equ{e3.5}) exhibits the convergence of the map
$(B,\b)\otto (\x,\g)$, since $\x<1$ in the region above the
separatrix: in the latter region Eq.(\equ{e3.5}) provides a convergent
expansion of the solution.

\def\SEC{Hyperbolic Coordinates}
\section{Hyperbolic Coordinates}\iniz

Motions with initial coordinate $\b(0)\ne0$ also admit a rather simple
representation. Remark that all pendulum motions with $\dot\b>0$
(hence different from the two equilibria) pass at some time through
a phase space point with $\b=0$. If $\dot\b$ is their velocity at
that moment we can find a quantity $\x$ such that $\dot\b,\b$ are
given by Eq.(\equ{e3.5}) with $\g=\d=1$ Therefore they can be
represented, at least as long as $U>0,\dot\b>0$ by introducing the {\it
dimensionless} variables $q'=\g\sqrt\x,p'=\d\sqrt\x$ and allowing
$\d,\g$ to be arbitrary.  Then the motions will be $t\to
(p'e^{g_0t},q'e^{-g_0t})$ showing that the motion can be represented
by the following two functions,

\be
\eqalign{
\txt S'=&\txt4\mathop\sum\limits_{m=0}^\infty
\big(\arctan((p'q')^m \,q')-\arctan((p'q')^m\,p')\big),
\cr
\txt R'=&\txt 4I\,g_0\mathop\sum\limits_{m=0}^\infty
\big(\frac{(p'q')^m p'}{1+((p'q')^{m} \,p')^2}+
\frac{(p'q')^m q'}{1+((p'q')^{m}\, q')^2}\big)\cr}
\Eq{e4.1}\ee
The motions $t\to (p'e^{g_0t},q'e^{-g_0t})$ solve the equations of
motion if $p',q'$ ({\it i.e.} $\g,\d$) are positive. But the equations
of motion are analytic, hence the formulae Eq.(\equ{e4.1}) together
with $t\to (p'e^{g_0t},q'e^{-g_0t})$, with $g_0=g_0(p'q')$, give
solutions of the pendulum equations independently of the sign of
$p',q'$, provided the series converge. The convergence requires
$|p'q'|<1$: which represents many data, in particular those in the
vicinity of the separatrix.

The coordinates can be called ``hyperbolic'' being suitable to
describe motions near the separatrix (where $p'q'=0$).  We also see
that time evolution preserves both volume elements $dBd\b$ and
$dp'dq'$; which means that the Jacobian determinant
$\frac{\partial(B,\b)}{\partial(p',q')}$ must be a function constant over the
trajectories, hence a function $D(x')$ of $x'\defi p'q'$. Note that
$D(x')$ has dimension of an action.

It is then possible to change coordinates setting $p =a(x') p',
q=a(x')q'$ and choose $a(x)$ so that the Jacobian determinant for
$(B,\b)\otto(p,q)$ is $\equiv1$.  A brief calculation shows that this
is achieved by fixing

\be a^2(x')=\frac1{x'}\int_0^{x'} D(y)dy,\Eq{e4.2}\ee
which is possible for $x$ small because, from Eq.(\equ{e3.5}) and (\equ{e2.2}),
it is $D(0)=32 I g>0$. Therefore the variables, which will have the
dimension of $a$, hence of a square root of an action,

\be p=p' \,{a(x')},\qquad  q=q' \,{a(x')},\Eq{e4.3}\ee
have Jacobian determinant $1$ with respect to $(B,\b)$ and the map
$(B,\b)\otto$ $(p,q)$ is area preserving, hence {\it canonical}.  The
Hamiltonian Eq.(\equ{e1.2}) becomes a function $\UU(x)$ of $x =pq$ and
the derivative of the energy with respect to $x$ has to be $g_0(x')$
(because the $p,q$ are canonically conjugated to $B,\b$).  Note that
$x$ has the dimension of an action, while $p,q$ are, dimensionally,
square roots of action.

This allows us to find $D(x')$: by imposing that the equations of
motion for the $(p,q)$ canonical variables have to be the Hamilton's
equations with Hamiltonian $\UU(x)\defi U(x')\equiv H(B, \b)$ it
follows that $\frac{d\UU(x)}{dx}=g_0(x')$, {\it i.e.}
$\frac{d\,U(x')}{dx'} \frac{dx'}{d x}=g_0(x')$ or
$\frac{d\,U(x')}{dx'}= g_0(x')(\frac{d}{dx'} (x'\,
a(x')^2))=g_0(x')D(x')$ by the above expression for $a(x')$. The just
obtained relation, together with Eq.(\equ{e2.2}), gives

\be D(x')=g_0(x')^{-1} \frac{d}{dx'}U(x')\Eq{e4.4}\ee
which is an explicit expression for the Jacobian
$\frac{\partial(B,\b)}{\partial(p',q')}\equiv
\frac{\partial(R,S)}{\partial(p',q')}=\frac{\partial(p,q)}{\partial(p',q')}$
(note that the Jacobian between $(B,\b)$ and $(p,q)$ is identically $1$
by construction). Eq.(\equ{e4.4}) is dimensionally correct because
$x'$ is dimensionless so that $U(x')$ has the correct dimension ({\it
i.e.} energy).

The function $U(x')$ is in Eq.(\equ{e2.7}) where
$k^2=\frac{h^{'2}}{h^2}$, by Eq.(\equ{e2.1}), is related to $x'=\x(h)$ by
Eq.(\equ{e2.2}), so that \cite[(8.197.3),(8.197.4)]{GR965},

\be U(x')=2 g^2 I\frac1{k^2}=2 g^2 I\frac{h^2}{h^{'2}}=32 I g^2 x'
\prod_{n=1}^\infty \Big(\frac{1+x^{'2n}}{1-x^{'(2n-1)}}\Big)^8\Eq{e4.5}\ee

To complete the determination of the canonical hyperbolic coordinates
it remains to find an expression for $D(x'), \UU(x)$ in terms of the elliptic
functions to obtain the canonical variable and the Hamiltonian in
closed form (rather than as power series as done so far).

\section{Determination of the Jacobian. Remarks}\iniz

It is remarkable is that the function $a^2$ defined above, hence such
that $D(x')=\frac{d}{dx'} (x'\,a^2(x'))$, seems to be simply

\be  a^2(z)=8 {I} \frac{ d}{dz}g_0(z),\Eq{e5.1}\ee
in a common holomorphy domain, for both sides, around $z=0$. This is
suggested by the agreement of the first $200$ coefficients of the
expansion of the two sides in powers of $z$: however this is not a
proof and the relation Eq.(\equ{e5.1}) holds because it can be seen to
be equivalent to an identity on elliptic functions, as discussed in
Appendix A below.  \*

\0{\it Remarks:} (1) The expansion of $D(x')$ in powers of $x'$ can be
derived from Eq.(\equ{e4.5}),(\equ{e2.6}), while that of $a^2(x')$ is
obtained from Eq.(\equ{e5.1}) and, again, Eq.(\equ{e2.6}).
\\
(2) It is perhaps natural to guess that the function
$a(x')^2$ should be closely related to $g_0(x')$; this is a guide to
its determination as it becomes, then, natural to look for it among
the derivatives of $g_0$ with respect to $x'$. By dimensional analysis
all $x'$-derivatives of $I g_0$ have the same dimension as
$a^2$. Looking {\it also} at the derivatives of $g_0$ as candidates
for $a^2$ is an idea due to one of us (PG). This follows a similar
line of thought which led to a conjecture on the canonical
integrability of the ``Calogero lattice'', \cite{GM973}, whose proof
was discovered in two subsequent works \cite{Mo975} and
\cite{Fr988}.\vskip2mm

The relation Eq.(\equ{e5.1}) is equivalent to a notable identity between
elliptic functions, as discussed in Appendix A below.
\*

Other peculiarities are, setting $32Ig=1,g=1$,
\*

\0(1) The function $g_0(x'),U(x')$, hence $\frac{d}{dx'}g_0(x'),
D(x')$, have Taylor coefficients in powers of $x'$ which {\it are} all
positive integers as it follows from the relations
Eq.(\equ{e4.5}) and Eq.(\equ{e2.6}), while $\UU(x)-\,x$ seems to have
alternating sign Taylor coefficients:

\be
\UU(x)-x=\,2x^2-4x^3+20x^4
-132 x^5+1008 x^6+\ldots\,\Eq{e5.2}\ee
where $\UU(x)$ is obtained by power series inversion of $x=x' a(x')^2$
and from $\UU(x)=U(x')$ together with Eq.(\equ{e4.5}).

\0(2) The function $U(x')$, energy of the pendulum expressed as a
    function of $x'$, has also the form

\be U(x')= 32 I g_0^2 \big[p' U_{x'}(p')+q'V_{x'}(q')\big]
\big[p' V_{x'}(p')+q'U_{x'}(q')\big]\defi x' \,f({x'})\Eq{e5.3}\ee
which, remarkably, has by Eq.(\equ{e4.5}) to depend only upon $x'$,
and have the form $x' f(x')$ for some $f$. This is not {\it a
priori} evident, unless the mechanical interpretation is kept in
mind, from the expressions found for $U,V$, namely

\be U_{x'}(z)=\sum_{\ell=0}^\infty
\frac{{x'}^{2\ell}}{1+(x'^{2\ell}z)^2},\qquad
V_{x'}(z)=\sum_{\ell=0}^\infty
\frac{{x'}^{2\ell+1}}{1+(x'^{2\ell+1}z)^2},\Eq{e5.4}\ee
\*

\0(3) Existence of an analytic canonical map integrating, near the
hyperbolic point, the system with energy Eq.(\equ{e1.1}) into one with
Hamiltonian $\UU(pq)=g\,pq+O((pq)^2)$ is well known: it
can be established without an explicit calculation by perturbation
analysis, see \cite[Appendix A3]{CG994}, for instance .

\appendix
\section{Proof of Eq.(\equ{e5.1})}
\iniz
\renewcommand{\theequation}{A\arabic{section}.\arabic{equation}}

Calling ${\EJ}(k)=\int_0^{\frac\p2}(1-k^2\sin^2\a)^{\frac12}d\a$ it is
$\EJ(h)=h{h'}^2 \frac{d \KJ(h)}{dh}+{h'}^2 \KJ(h)$, see
\cite[(8.123.2]{GR965}, and
$\EJ(h)\KJ(h')+\EJ(h')\KJ(h)-\KJ(h)\KJ(h')=\frac\p2$, see
\cite[(8.122]{GR965}; the latter ``Legendre's relation'',
\cite[p.520]{WW927}, combined with $\frac{dh'}{dh}=-\frac{h}{h'}$
yields the identity

\be h' h^2\Big(\KJ(h)\frac{d \KJ(h')}{d h'}- \KJ(h')\frac{d \KJ(h)}{d
h'}\Big)= \frac\p2\,.\Eq{eA1.1}\ee
This can be used to obtain an expression for $\frac{d\log x'}{dh}$:
keeping in mind $x'=e^{-\p \KJ(h')/\KJ(h)}$ it is
$\frac{d \log x'}{dh'}=-\p \big(\frac{1}{\KJ(h)}\frac{d
\KJ(h')}{dh'}-\frac{\KJ(h')}{\KJ(h)^2} \frac{d\KJ(h)}{dh'}\big)$ which
is transformed into
$\frac{d \log x'}{dh'}= \log x'\,
\big(\frac1{\KJ(h')}\frac{d\KJ(h')}{dh'}
-\frac1{\KJ(h)}\frac{d\KJ(h)}{dh'}\big)$.

Form Eq.(\equ{eA1.1}) it follows, therefore,

\be \frac{d}{d h'} \log x'=\frac{\p}2\frac{\log
  x'}{h'h^2\KJ(h)\KJ(h')}\qquad
\frac{d}{d h} \log x'=-\frac{\p}2\frac{\log
  x'}{h{{h'}^2}\KJ(h)\KJ(h')}\,.\Eq{eA1.2}\ee
and the corresponding derivatives with respect to $h$ are obtained by
multiplying both sides by $-\frac{h}{h'}$.

To establish Eq.(\equ{e5.1}) consider the relation,

\be
\frac{d}{dh}\Big\{h {h'}^{2}\,\frac{d
  \KJ(h)}{dh}\Big\}\Eq{eA1.3}-h \KJ(h)=0\ee
see \cite[(8.124.1]{GR965}. This implies by simple algebra, and
keeping in mind that $\frac{h}{h'}=-\frac{d h'}{dh}$,  the
following identity

\be h\,\KJ(h)= {h'}^3\,\frac{d}{dh} \Big(h {h'}^2\big(-\frac1{h'}
\frac{d}{dh}\KJ(h) +\frac{h}{{h'}^3}\KJ(h)\big)\Big)\Eq{eA1.4}\ee
which is a known linear equation, solved by $\KJ(h)$. This can be
rewritten, since $\frac{h}{h'}=-\frac{d h'}{dh}$, as

\be \eqalign{
&\frac{h}{{h'}^3}\,\KJ(h)=
\frac{d}{dh} \Big(h {h'}^2\big(-\frac1{h'}
\frac{d}{dh}\KJ(h) -\frac1{{h'}^2}\frac{d h'}{dh}
\KJ(h)\big)\Big)\cr
&=\frac{d}{dh} \Big(h {h'}^2\KJ(h)^2
\frac{d}{dh}\frac1{h'\KJ(h)}\Big)\cr
}\Eq{eA1.5}\ee
Remarking that $\frac{2h}{h'^4}\equiv \frac{d}{dh}\frac{h^2}{{h'}^2}$,
Eq.(\equ{eA1.5}) implies, multiplying both sides by $\frac{2}{{h'\KJ(h)}}$,

\be
\eqalign{
&\frac{d}{dh}\big(\frac{h}{h'}\big)^2=\frac{2}{h'\KJ(h)}\frac{d}{dh}\Big(h
h'^2 \KJ(h)^2\big(\frac{d}{dh}\frac1{h'\KJ(h)}\big)\Big)\cr
&=
\frac{2\p}{h'\KJ(h)}\frac{d}{dh}
\Big(  \frac{h
h'^2 \KJ(h)\KJ(h')}{\p\KJ(h')/\KJ(h)}
\big(\frac{d}{dh}\frac1{h'\KJ(h)}\big)\Big)\cr}
\Eq{eA1.6}\ee
and by the first of Eq.(\equ{eA1.2}) multiplied by
$\frac{dh'}{dh}=-\frac{h}{h'}$
this is, using $k^2=\frac{h^2}{{h'}^2}$

\be \eqalign{
&\frac{d}{dh}\frac1{k^2}= \frac{\p^2}{h'\KJ(h)}
\frac{d}{dh}\Big(\frac{dh}{d\log x'}
\big(\frac{d}{dh}\frac1{h'\KJ(h)}\big)\Big)\cr
&= \frac{\p^2}{h'\KJ(h)}
\frac{d}{dh}\Big(x'\frac{d}{dx'} \frac1{h'\KJ(h)}\Big)
\cr}
\Eq{eA1.7}\ee
and multiplying by $2Ig^2\frac{dh}{dx'}$ it follows

\be 2Ig^2\frac{d}{dx'} \frac1{k^2}=
8I \frac{\p g}2 \frac1{h'\KJ(h)}\frac{d}{dx'}\big(x'\frac{d}{dx'}
\frac{\p g}{2} \frac1{h'\KJ(h)} \big)\Eq{eA1.8}\ee
and setting $a(x')^2\defi 8I \frac{d}{dx'} \frac{\p g}{2}
\frac1{h'\KJ(h)}\equiv 8I \frac{d}{dx'}g_0(x')$ the last relation is\\
$\frac{d}{dx'} U(x')= g_0(x') \frac{d}{dx'} \big(x'a(x')^2\big)$ so
that Eq.(\equ{e4.4}) and Eq.(\equ{e4.2}) imply Eq.(\equ{e5.1}).

\section{Pendulum at the stable equilibrium}
\iniz
\renewcommand{\theequation}{A\arabic{section}.\arabic{equation}}

From the above results it is straightforward to find the canonical
transformation that converts the pendulum Hamiltonian in its normal
form around the stable equilibrium point. The Hamiltonian is now given
by Eq.(\equ{e1.1}) with the substitution: $g=ig_s$. It is natural to
define $k_s=ik$ in order to use the same set of equations from the
unstable case. The system energy is then $U_s=2g_s^2/k_s^2$ and large
values of $k_s$ correspond now to small oscillations around the
equilibrium point.

Finally, it is convenient to define
\be
\eqalign{
k_s'\,=\,&\sqrt{1-k_s^2},\qquad
h_s'(k_s)\,=\,\frac{k_s}{\sqrt{k_s^2-1}},\qquad {h_s}\,=\,\sqrt{1-{h_s'}^2}
\cr
h'(k)\,=\,&\frac{1}{h_s'(k_s)},\qquad
h(k)\,=\,\frac{i h_s(k_s)}{h_s'(k_s)}
\cr}\Eq{eA2.1}\ee
and one finds:
\be
\eqalign{
g_0^{(s)}(h_s)\,=\,&-i g_0(h)\,=\,\frac{\p}{2}\frac{g_s}{\KJ(h_s)}\cr
x_s'(h_s)\,=\,&e^{-\p\KJ(h_s')/\KJ(h_s)}=-x'(h)
\cr}\Eq{eA2.2}\ee
where we have used \cite[(8.128)]{GR965}.

With these conventions, and going through computations similar to the
ones performed to study the unstable point, the relations found for the
latter can be converted into the corresponding ones for the
equilibrium point. In particular, by choosing
$p'=\sqrt{x'_s}\cos(g_0^{(s)}t)$ and $q'=\sqrt{x'_s}\sin(g_0^{(s)}t)$
the transformation given by Eq.(\equ{e4.1}) is now:

\be
\eqalign{
\txt S_s'=&\txt{4}i\mathop\sum\limits_{m=0}^\infty (-1)^m
\big(\arctan((p'^2+q'^2)^m \,(p'+iq'))\cr
&-\arctan((p'^2+q'^2)^m\,(p'-iq'))\big),
\cr
\txt R_s'=&-\txt 4I\,g_0^{(s)}\mathop\sum\limits_{m=0}^\infty (-1)^m
\big(\frac{(p'^2+q'^2)^m (p'+iq')}{1-((p'^2+q'^2)^{m} \,(p'+iq'))^2}
\cr
&+
\frac{(p'^2+q'^2)^m (p'-iq')}{1-((p'^2+q'^2)^{m}\, (p'-iq'))^2}\big)\cr}
\Eq{eA2.3}\ee
where the relation
$R_s'=g_0^{(s)}I(p'\partial_{q'}-q'\partial_{p'})S_s'$ holds. And the energy 
can be written (see Eq.(\equ{e4.5})): 

\be U_s(x_s')=2 g_s^2 I\frac1{k^2}=32 I g_s^2 x_s'
\prod_{n=1}^\infty \Big(\frac{1+x_s^{'2n}}{1+x_s^{'(2n-1)}}\Big)^8\Eq{eA2.4}\ee

The
transformation $(B,\beta)\rightarrow (p'.q')$ is not canonical. The canonical variables, $(p,q)$, can
be found by looking for a function $a_s(x_s')$ (which depends on the
constant of motion $x_s'$) such that $(p,q)=(a_s(x_s')p',a_s(x_s')q')$
and the Jacobian of the transformation is one. It is, as in the
hyperbolic case,

\be  a_s^2(z)=-16 {I} \frac{ d}{dz}g_0^{(s)}(z),\Eq{eA2.5}\ee

Finally, the normal form of the Hamiltonian now reads:
\be
\UU_s(x)=32Ig_s^2 \,W(\frac{x}{64Ig_s})\ee
where
\be
W(z)=z\,(1-2z-4z^2-20z^3-132z^4-1008z^5\ldots )
\ee
which can be compared with the hyperbolic case expression Eq.(\equ{e5.2}):
\be
\UU(x)=-32Ig^2\,W(-\frac{x}{32Ig})
\ee

\small
\bibliographystyle{unsrt}

\end{document}